\documentclass[twoside]{photon2007}
\usepackage[latin1]{inputenc}
\usepackage[dvips]{graphicx,epsfig,color}
\usepackage{wrapfig,rotating}
\usepackage{amssymb,amsmath,array}

\begin{document}
\title{
%%%%   Paper title goes here  %%%%%%%%%%%%%%
Ultraperipheral Collisions at RHIC and LHC } 
%% 
%***********************************************************************
% AUTHORS INFORMATION AREA
%***********************************************************************
\author{Gerhard Baur
%$^1$
% Optional short acknowledgment: remove next line if non-needed
%\thanks{This is an optional funding source acknowledgment.}
% DO NOT MODIFY THE FOLLOWING '\vspace' ARGUMENT
\vspace{.3cm}\\
% Addresses and institutions (remove "1- " in case of a single institution)
Institut f\"ur Kernphysik,
Forschungszentrum J\"ulich,
D-52425 J\"ulich, Germany 
%% Remove the next three lines in case of a single institution
%\vspace{.1cm}\\
%2-  Second Author's Institute \\
%Address of Second Author's Institute\\
}
%%***********************************************************************
% END OF AUTHORS INFORMATION AREA
%***********************************************************************

\maketitle

\begin{abstract}
 A brief introduction to the physics of ultraperipheral
collisions at collider energies is given. Photon-hadron
(proton/ nucleus) and photon-photon
interactions can be studied in a hitherto unexplored energy 
regime.
\end{abstract}

\section{Some features of Ultraperipheral Collisions
(UPC)}
Photon-photon and photon-hadron interactions
can also be studied in hadron-hadron collisions~\cite{url}.
This may be surprising since in general such collisions are
dominated by strong interactions between the hadrons.
However, by choosing collisions with large 
impact parameter b (or, equivalently, small momentum transfer)
one can suppress these strong interactions.

\begin{figure}[h]
\centerline{\includegraphics[width=1.0\columnwidth]{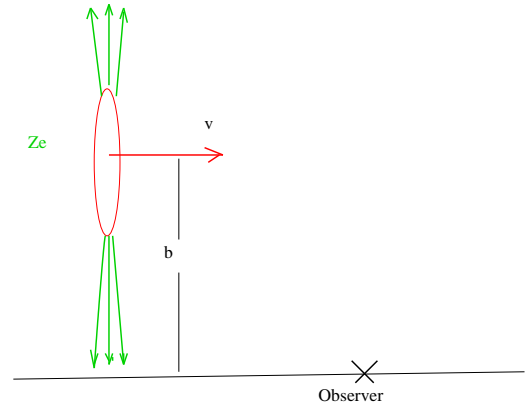}}
\caption{A fast charged particle 
moving on a straight line with impact parameter b causes a time-dependent
electromagnetic field at the point of the observer.
This field corresponds to a spectrum of equivalent photons.}\label{Fig:PH}
\end{figure}

The time-dependent electromagnetic field of a fast moving charged particle
can be thought of as a spectrum of (quasireal, or equivalent)
photons~\cite{fermi}, see Figure \ref{Fig:PH}.
The determination of the 
equivalent (or Weizs\"acker-Williams) photon spectrum
corresponding to a fast particle moving past an observer on a 
straight line path with impact parameter $b$ is a textbook example
~\cite{jac}. 

The probability $P(b)$ of a specific
photon-hadron reaction to occur in a collision
with an impact parameter $b$
is given by $P(b)=N(\omega, b) \sigma_{\gamma h}(\omega)$,
where $\sigma_{\gamma h}$ is the corresponding photoproduction
cross section. The equivalent photon spectrum can be calculated
analytically, a useful approximation for qualitative considerations is 
\begin{equation}
N(\omega, b)=\frac{Z^2 \alpha}{\pi^2 b^2}
\label{Eq:nb}
\end{equation}
for $\omega<\frac{\gamma}{b}$
and zero otherwise. The nuclear charge is given 
by $Z$, heavy ions have particularly high photon fluxes,
however, this is partially offset by the lower ion-ion luminosities,
as compared to the p-p case.

%\begin{wrapfigure}{r}{1.0\columnwidth}
%\end{wrapfigure}
The impact parameter b is restricted to
\begin{equation}
b>b_{min} \sim R_1 + R_2
\end{equation}
where $R_1$ and $R_2$ denote the sizes of the hadrons.

For heavy ion scattering the Coulomb parameter
$\eta \equiv \frac{Z^2 e^2}{\hbar v}\sim Z^2/137$ is 
much larger than unity and it is in principle possible to determine the
impact parameter
by measuring the angle of Coulomb scattering.
Whereas this is experimentally feasible at lower 
($\sim GeV/A$) energies
~\cite{aum}, this angle is too small at collider
energies. So one generally measures quantities
integrated over all impact parameters.
Too small impact parameters are recognized since the event is
dominated by the violent strong interactions.

The photon spectrum Eq.~\ref{Eq:nb} extends up to 
a maximum photon energy given by 
\begin{equation}
\omega_{max}=\frac{\gamma}{b_{min}} .
\end{equation}
This energy is about 3 GeV at RHIC (Au-Au, $\gamma \sim 100$),
and 100 GeV at LHC (Pb-Pb, $\gamma \sim 3000$) in the collider
system. 

\section{Multiphoton processes: a possible trigger on UPC}
%% section headers !

\begin{figure}[h]
\centerline{\includegraphics[width=1.0\columnwidth]
{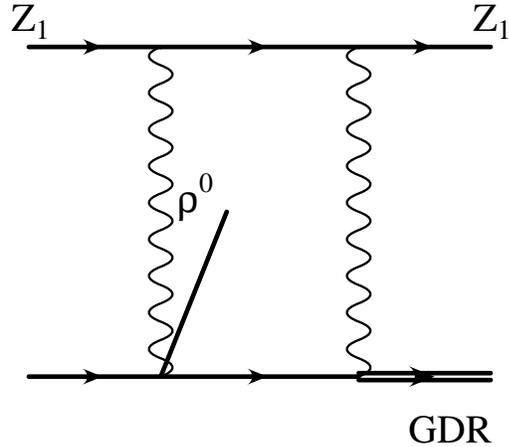}}
\caption{A graph contributing to the simultaneous 
production of a $\rho$-meson and the excitation of the 
giant dipole resonance (GDR).}\label{Fig:mua}
\end{figure}

For heavy ions the probability of an electromagnetic
interaction in ultraperipheral collisions is especially 
large, and multiphoton processes occur, see e.g.~\cite{npa}.
We mention $e^+e^-$ pair production where 
the impact parameter dependent total pair
production probability $P(b)$
is of order unity. Multiple pairs can be produced,
however they may be hard to detect due to their low
transverse momentum.
The nuclear giant dipole resonance is excited with  
probabilities of order of one third.
In Figure \ref{Fig:mua} one of the
graphs is shown which leads to the electromagnetic 
production of a $\rho^0$ along with the excitation 
of the giant dipole resonance. 
These graphs can conveniently be evaluated in semiclassical or eikonal
theories~\cite{npa}.

%\begin{figure}[h]
%\centerline{\includegraphics[width=1.0\columnwidth]{fig3b.ps}}
%\caption{A fast (relativistic charged particle...}\label{Fig:mub}
%\end{figure}
%Captions of figures and tables appear {\em below} the figure/table.
%When referring to Figure~\ref{Fig:PH} capitalize the first letter.
The giant dipole resonance decays dominantly into a neutron.
This neutron is detected in the forward direction and can serve as 
a trigger on UPC.

\section{UPC at RHIC}
The physics of UPC at RHIC and 
results from the STAR detector were covered by J. Seger
in the session on photon- and electroweak boson physics,
from HERA, RHIC and Tevatron to LHC.

A unique feature to photoproduction 
in hadron-hadron collisions is an interference effect
~\cite{kn}: a vector meson can be produced by a photon
originating from either of the hadrons. 
It was shown in~\cite{kn} that this interference effect 
leads to a reduction of the transverse momentum spectrum
of the vector mesons for small transverse momenta.
Another theoretical approach~\cite{hbt} 
leads to very similar conclusions. (Preliminary) 
experimental results from STAR/RHIC indeed show a dip
for small transverse momenta, see e.g. Ref. \cite{yr}. 

\section{Opportunities for UPC at LHC}
The maximum photon energy scales linearly with 
the Lorentz factor $\gamma$, see eq. 3. This leads to
a significant widening of the opportunities at LHC as compared to
RHIC. A most promising area is low-x QCD studies.
The experiments at HERA have shown that 
photoproduction processes provide a well-understood
probe of the gluon density in the proton. At LHC,
such processes could be extended to invariant
$\gamma p$ energies exceeding the maximal HERA energy
by a factor of 10. This would allow to use dijet (charm, etc.)
production to measure the gluon density in the proton
and/or nucleus down to $x \sim 3 \times 10^{-5}$.
 Ultraperipheral collisions would also allow one 
to study the coherent production of heavy quarkonia,
$\gamma + A \rightarrow J/\Psi (\Upsilon) + A$
at $x \lessapprox 10^{-2}$, and to investigate the 
propagation of small dipoles through the nuclear medium
at high energies, see Ref. \cite{fra}, see also  
Refs.~\cite{kopel,macha}. 
Dijet production via photon-gluon fusion is 
calculated in Ref. \cite{svw}. Very large rates 
are obtained that will considerably extend the HERA x range.

In this session plans for 
studying UPC physics with heavy ions at the LHC were covered 
by J.Nystrand (ALICE), D.D'Enterria (CMS),
and V. Pozdnyakov (ATLAS). 

In addition to diffractive processes in proton-proton 
collisions at LHC also a rich program of proton-photon 
and photon-photon physics can be pursued, see Ref. \cite{cern}.
The photon flux is lower as compared to the 
heavy ion case due to the $Z^2$-factor, but this
is at least partly compensated by higher beam luminosities.
The photon spectrum is harder due to the smaller size as compared to 
the heavy ions, this leads  to a lower value of $b_{min}$ in Eq. 3 . 
Possibilities for electroweak physics and
beyond were presented by S.Ovyn ($\gamma p$) and T. Pierzchala
($\gamma \gamma$) in this session.
Tagging on photon energy by measuring the energy loss of 
the scattered protons in the forward detector TOTEM
is an important feature.
 In this session J.Pinfold reported on
photon-photon, photon-pomeron and double pomeron
production at CDF.

A recent workshop on photoproduction at collider energies 
at ECT*/Trento was devoted to UPC, the mini-proceedings 
can be found in~\cite{ect}. 
The reviews~\cite{ber,soff,bau,bns} and the 
most recent preprint~\cite{yr} reflect the gradual progess 
of the field. 
   
\section*{Acknowledgments}

I would like to thank Frederic Kapusta
for his kind invitation to this very pleasant 
and interesting conference
at such a venerable place.

% ****************************************************************************
% BIBLIOGRAPHY AREA
% ****************************************************************************

\begin{footnotesize}
% IF YOU DO NOT USE BIBTEX, USE THE FOLLOWING SAMPLE SCHEME FOR THE REFERENCES
% ----------------------------------------------------------------------------

% ----------------------------------------------------------------------------

% IF YOU USE BIBTEX,
% - DELETE THE TEXT BETWEEN THE TWO ABOVE DASHED LINES
% - UNCOMMENT THE NEXT TWO LINES AND REPLACE 'Name_Of_Your_BibFile'

%\bibliographystyle{unsrt}
%\bibliography{Name_Of_Your_BibFile}

\begin{thebibliography}{99}
% Please replace the numbers for   contribId   and   sessionId
% in the following URL. You can get this information by going to 
% http://indico.cern.ch/confAuthorIndex.py?confId=9499
% and search for your contribution and click on the title
% Be aware: '&amp;' must be replaced by simple '&' as in example below
\bibitem{url} Slides: \\ 
\verb$http://indico.cern.ch/materialDisplay.py?$ \\
\verb$contribId=39&sessionId=16$ \\
\verb$&materialId=slides&confId=3841$
%------- replace following references ;-)
%\bibitem{parton_qed} A.D.~Martin {\it et~al.}, Eur. Phys. J. {\bf C39} 155 (2005).
%\bibitem{H1}N.~Gogitidze, arXiv:hep-ex/0701033 (2007).
%\bibitem{DVCS}S.~Friot, B.~Pire and L.~Szymanowski, Phys. Lett. {\bf B645} 153 (2007);\\
%              D.~Hasell, R.~Milner and K.~Takase, 
%AIP Conf. Proc. {\bf 588} 187 (2001);\\
%              M.~Krawczyk and A.~Zembrzuski, Phys. Rev. {\bf D57} 10 (1998).
%\bibitem{pomeron}R.~Brower and C.~Tan, PoS LAT2005 279 (2006);\\
%                 J.P.~Guillaud and A.~Sobol,
%  {\it Perspectives of the study of double Pomeron exchange at the LHC},
%  11th Lomonosov Conference on Elementary Particle Physics, Moscow, Russia (2003).
\bibitem{fermi}E.~Fermi, Z.Physik {\bf 29} 315 (1924)
\bibitem{jac} J.D. Jackson, Classical Electrodynamics (Wiley, New York,1975)
\bibitem{aum} T.~Aumann, Eur. Phys. J. {\bf A26} 441 (2005)
\bibitem{npa} G.~Baur {\it et~al.}, Nucl. Phys. {\bf A729} 787 (2003)
\bibitem{kn} Spencer~R.~Klein and Joakim~Nystrand Phys. Rev. Lett., {\bf 84} 
2330 (2000) 
\bibitem{hbt} K.~Hencken, G.~Baur, and D.~Trautmann, Phys. Rev. Lett.
{\bf 96} 012303 (2006)
\bibitem{yr} A.~Baltz {\it et~al.}, arXiv:0706.3356
\bibitem{fra} Leonid Frankfurt, Mark Strikman, and Christian Weiss
 Annual Review of  Nuclear and Particle Science
{\bf 55} 403 (2005) 
\bibitem{kopel} Yuri Ivanov, Boris Kopeliovich, and
Ivan Schmidt, arXiv:0706.1532
\bibitem{macha} V.~P.~Gon\c{c}alves and M.~V.~T.~Machado, arXiv:0706.2810
\bibitem{svw} Mark Strikman, Ramona Vogt, Sebastian White
Phys. Rev. Lett. {\bf 96} 082001 (2006)
\bibitem{cern} Prospects for Diffractive and Forward Physics at the
LHC CERN/LHCC 2006-039/G-124
\bibitem{ect} A.~Baltz {\it et~al.}, arXiv: hep-ph/0702212
\bibitem{ber} C.~A.~Bertulani, and G.~Baur, Phys. Rept. {\bf 163} 299 (1988)
\bibitem{soff} F.~Krauss, M.~Greiner, G.~Soff, Prog. Part. Nucl. Phys. 
{\bf 39} 503 (1997)
\bibitem{bau} G.~Baur {\it et~al.}, Phys. Rept. {\bf 364} 359 (2002)
\bibitem{bns} Carlos~A.~Bertulani, Spencer~R.~Klein,
and Joakim Nystrand, Annual Review of  Nuclear and Particle Science
{\bf 55} 271 (2005)  
\end{thebibliography}
% example of Name_Of_Your_BibFile.bib
% @Article{Turcato:2006ch,
%      author    = "Turcato, M.",
%  collaboration = "ZEUS and H1",
%      title     = "Lepton flavour violation and charmonium physics at HERA",
%      journal   = "Nucl. Phys. Proc. Suppl.",
%      volume    = "162",
%      year      = "2006", 
%      pages     = "283-287",
%      SLACcitation  = "%%CITATION = NUPHZ,162,283;%%"
% }
% 
% @Unpublished{Gogitidze:2007du,
%      author    = "Gogitidze, N.",
%  collaboration = "H1", 
%      title     = "Prompt photons and particle momentum distributions at
%                   HERA", 
%      year      = "2007",
%      note    = "hep-ex/0701033",
%      SLACcitation  = "%%CITATION = HEP-EX 0701033;%%"
% }

\end{footnotesize}

% ****************************************************************************
% END OF BIBLIOGRAPHY AREA
% ****************************************************************************

\end{document}